\documentstyle[12pt,epsf]{article}
\newcommand{\lsm}{L$\sigma$M}
\newcommand{\fsigma}{\mbox{$f_{0}$(400--1200)}}
\newcommand{\fn}{\mbox{$f_{0}$(980)}}
\newcommand{\ft}{\mbox{$f_{0}$(1370)}}

\begin{document}\baselineskip .7cm
\title{\bf Remarks on the \fsigma\ scalar meson as the dynamically
generated chiral partner of the pion}
\author{Eef van Beveren$^{\,a}\!\!$\,,
Frieder Kleefeld$^{\,b}\!\!$\,,
George Rupp$^{\,b}\!\!$\,, and
Michael D.\ Scadron$^{\,c}\!\!$ \\[5mm]
$^{a}${\footnotesize\it Departamento de F\'{\i}sica, Universidade de Coimbra,
P-3004-516 Coimbra, Portugal} \\ {\footnotesize\tt eef@teor.fis.uc.pt}\\[.3cm]
$^{b}${\footnotesize\it Centro de F\'{\i}sica das Interac\c{c}\~{o}es
Fundamentais, Instituto Superior T\'{e}cnico, P-1049-001 Lisboa, Portugal} \\
{\footnotesize{\tt kleefeld@cfif.ist.utl.pt,}
{\tt george@ajax.ist.utl.pt} (corresponding author)} \\[.3cm]
$^{c}${\footnotesize\it Physics Department, University of Arizona, Tucson,
AZ 85721, USA} \\ {\footnotesize\tt scadron@physics.arizona.edu} \\[5mm]
[.3cm]
{\small hep-ph/0204139}
}
\date{\today}
\maketitle

\begin{abstract}
The quark-level linear $\sigma$ model (\lsm) is revisited, in
particular concerning the identification of the $f_0$(400--1200) (or
$\sigma$(600)) scalar meson as the chiral partner of the pion. We demonstrate
the predictive power of the \lsm\ through the $\pi\pi$ and $\pi N$ $s$-wave
scattering lengths, as well as several electromagnetic, weak, and strong
decays of pseudoscalar and vector mesons. The ease with which the data for
these observables are reproduced in the \lsm\ lends credit to the necessity to
include the $\sigma$ as a fundamental $q\bar{q}$ degree of freedom, to be
contrasted with approaches like chiral perturbation theory or the confining
NJL model of Shakin and Wang.
\end{abstract}
\section{Introduction}
The question as to whether the pion has a scalar $q\bar{q}$ partner remains
highly topical, now that the \fsigma\ (or $\sigma$) meson has become a firmly
established resonance \cite{PDG00}. For the latter reason, the bone of
contention has shifted from the cavilling at the ``existence'' of the
$\sigma$ towards a somewhat more sensible discussion whether the $\sigma$ is a
``fundamental'' or a ``dynamically generated'' particle. Now while
there is little dispute about what ``fundamental'' (or ``intrinsic'') means in
a mesonic context, namely a totally colorless state composed of normally one
but possibly more $q\bar{q}$ pairs, the term ``dynamically generated'' (or
``dynamical'' only) has been used by several authors to express rather
different physical mechanisms.

For instance, in a Comment \cite{IS97} on a paper by T\"{o}rnqvist and Roos
(TR) \cite{TR96}, Isgur and Speth (IS) argued that the $\sigma$ meson, at least
in their approach, is a broad ``dynamical pole'' due to $t$-channel forces
only, arising from degrees of freedom already present in the meson-meson
continuum, in contrast to an ``intrinsic pole'' resulting from a new $q\bar{q}$
degree of freedom in the dynamics. Moreover, IS criticized and drew into
question the conclusions of TR because of the omission of $t$-channel forces in
their work.  However, in another Comment \cite{HSS97} on TR's paper,
Harada, Sannino, and Schechter demonstrated in a concrete model calculation
that this omission appears to be not very crucial and only mildly affects the
$\sigma$-meson mass
and width.  Also in the unitarized meson model of two of us \cite{BR86_99}, the
$\sigma$ resonance is a consequence of the inclusion of $p$-wave $q\bar{q}$
states, but strongly coupled to the meson-meson continuum via the $^3P_0$
mechanism. This gives rise to a doubling of the number of poles originally
present in the ground-state confinement spectra, the lower poles corresponding
to the light scalar mesons like the $\sigma$. In this formalism, it makes
little sense to talk about ``intrinsic'' versus ``dynamical'' poles, since the
whole unitarization scheme is highly dynamical, producing large effects that
strongly influence all poles. A similar conclusion has been reached very
recently by Boglione and Pennington \cite{BP02}.

In chiral-symmetric approaches like the quark-level Linear $\sigma$ Model
(\lsm) \cite{GML60,DS95_99} and the Nambu--Jona-Lasinio (NJL) model
\cite{NJL61}, the $\sigma$ meson naturally appears as the chiral scalar
$q\bar{q}$ partner of the pion. Moreover, in the \lsm\ the $\sigma$, which is
introduced as an elementary degree of freedom in the Lagrangian, is also
self-consistently generated in loop order through a quark loop and tadpole
\cite{DS95_99}. So the $\sigma$ meson is \em both \em \/``fundamental'' \em
and \em \/``dynamically generated''. On the other hand, in the \em confining
\em \/NJL model of Shakin and Wang (SW) \cite{SW_prd64,SW_prd63}, no light
scalar $q\bar{q}$ state shows up, in contrast to the traditional NJL approach.
However, SW do
predict a light scalar resonance, which could  be interpreted as the \fsigma,
merely through $t$- and $u$-channel $\rho$ exchange in $\pi\pi$ scattering
\cite{SW_prd63}, in much the same way as IS \cite{IS97} (see above). Such
states SW call ``dynamically generated'' resonances, as opposed to
``pre-existing'' ones. These model results have led them to conclude that
\cite{SW_prd64} \em ``the $\sigma$ obtained from the study of $\pi\pi$
scattering is not the chiral partner of the pion'' \em \/and \em ``the
non-linear sigma model is the model of choice''. \em In Ref.~\cite{SW_prd63},
SW also arrive at several other conclusions on the nature of different scalar
mesons, which we have shown \cite{RBS01} to be not supported by experiment
(see also Ref.~\cite{BRSK00_01}). In Ref.~\cite{RBS01}, we also argued against
the strict distinction between ``intrinsic'' and ``dynamically generated''
scalar-meson states made by SW.

In the present paper, we readdress the issue of the pion's chiral $q\bar{q}$
partner, and reach the following conclusions:
\begin{enumerate}
\item indeed an $f_0$(630) is dynamically generated from the chiral field
theory constituted by the quark-level linear $\sigma$ model (L$\sigma$M) 
\cite{DS95_99}, which is based on the original chiral Gell-Mann--L\'{e}vy
nucleon L$\sigma$M \cite{GML60}, but also predicts the famous NJL \cite{NJL61}
result $m_{\sigma}=2m_q$; 
\item the above SW conclusion on the nature of the $\sigma$ meson is incorrect.
Instead, this $\sigma$(630) is indeed the scalar $n\bar{n}$ chiral partner of
the pion. 
\end{enumerate}
Rather than just repeating the analysis of Ref.~\cite{DS95_99}, in Sec.~2 we
demonstrate the chiral structure of these \lsm\ states for strong,
electromagnetic (e.m.), and weak interactions. In Sec.~3, we verify our use of
chiral \lsm\ couplings by showing that corresponding P$\gamma\gamma$,
VP$\gamma$ plus PV$\gamma$, VPP e.m., and strong \lsm\ quark loop-order graphs
are always compatible with observed \cite{PDG00} $SU(2)$ and $SU(3)$ sum-rule
data. We draw our conclusions in Sec.~4 and, in passing, note that both the
$SU(3)$-symmetry and infinite-momentum-frame approaches of Ref.~\cite{S82_92},
and also the dynamical unitarized nonet scheme of Ref.\cite{BR86_99}, arrive at
\em different \em \/$q\bar{q}$ patterns for the isoscalar scalar mesons than
SW in Refs.~\cite{SW_prd64,SW_prd63}. 
\section{Why the $f_0$(630) scalar $\sigma$ meson is the chiral partner of the
$\pi$}
\subsection{Brief summary of the \lsm\ field theory}
The chiral-symmetric $SU(2)$ linear $\sigma$ model (\lsm) was first
formulated in 1960 \cite{GML60}, while the $SU(3)$ version dates from 1967,
1969 and 1971, respectively \cite{LGS67_71}. The \lsm\ pseudoscalar and scalar nonet
$U(3)$ states [$\pi$(140), $K$(492), $\eta$(549), $\eta^{\prime}$(958), and
$\sigma$(650), $\kappa$(800--900), \fn, $a_0$(980)] were later dynamically
generated \cite{S82_92,DS95_99}. A \lsm\ is manifestly renomalizable and much
easier to
handle than the non-linear NJL scheme \cite{NJL61}, yet chiral symmetry in fact
blends together these two pictures \cite{E76_78}, as the dynamically generated
theory \cite{DS95_99} shows. Specifically, the $SU(2)$ \lsm\ interaction
Lagrangian --- due to dynamical symmetry breaking \cite{DS95_99} or spontaneous
symmetry breaking \cite{S82_92} --- reads, after the shift of the $\sigma$
field,
\begin{equation}
{\cal L}^{\mbox{\scriptsize int}}_{\mbox{\scriptsize\lsm}} = g\,\bar{\psi}\,
(\sigma+i\gamma_5\vec{\tau}\cdot\vec{\pi})\,\psi\,+\,g'\,\sigma(\sigma^2+\pi^2)
\,-\,\frac{\lambda}{4}(\sigma^2+\pi^2)^2 \; .
\label{lsm}
\end{equation}
Here, the fermion fields refer to quarks \cite{DS95_99} and not to nucleons,
with constituent quark mass $m_q=\frac{1}{2}(m_u+m_d)$ generated via the chiral
Goldberger--Treiman relation (GTR) $f_{\pi}g=m_q$, with $f_{\pi}\approx$ 93 MeV
(and 90 MeV in the chiral limit (CL) \cite{CS81}), resulting in a value near
$m_q\approx m_N/3 \approx$ 315~MeV. In fact, it is dynamically generated in the
CL as $m_q\approx325$ MeV \cite{DS95_99}, and the Gell-Mann--L\'{e}vy chiral
relations at tree level are \cite{GML60}
\begin{equation}
g \; = \; \frac{m_q}{f_{\pi}} \;\;\;\; , \;\;\;\; g' \; = \;
\frac{m_{\sigma}^2}{2f_{\pi}} \; = \; \lambda f_{\pi} \; .
\end{equation}
Moreover, at one-loop level Eqs.~(2.2) are recovered, together with two new
equations \cite{DS95_99} in the CL:
\begin{equation}
m_{\sigma} \; = \; 2\,m_q \;\;\;\; , \;\;\;\; g_{\pi qq} \; = \; g \; = \;
\frac{2\pi}{\sqrt{N_c}} \; ,
\end{equation}
for $N_c=3$, also dynamically generated. Then, $g=2\pi/\sqrt{3}=3.6276$, and
\begin{equation}
m_{q}\;=\;2\pi\,\frac{f_{\pi}^{\mbox{\scriptsize CL}}}{\sqrt{3}}\;\approx
325\;\mbox{MeV} \;\;\; , \;\;\; m_{\sigma}\;=\;2m_q\;\approx\;650\;\mbox{MeV}
\end{equation}
are dynamically generated, from the chiral GTR \cite{DS95_99}. Finally, all
three \lsm\ couplings in Eq.~(2.1) are dynamically generated as
\begin{equation}
g\;=\;\frac{2\pi}{\sqrt{3}}\;\approx\;3.6\;\;\;,\;\;\;g'\;=\;2g\,m_q\;\approx\;
2.3\;\mbox{GeV}\;\;\;,\;\;\;\lambda\;=\;\frac{8\pi^2}{3}\;=\;26.3 \;.
\end{equation}

Furthermore, this \lsm\ then also recovers the vector-meson-dominance (VMD)
prediction $g_{\rho\pi\pi}=g_{\rho}$ from quark loops alone. When the
$\pi$-$\sigma$-$\pi$ \lsm\ meson loop is added, this VMD prediction is extended
to \cite{BRS98}
\begin{equation}
\frac{g_{\rho\pi\pi}}{g_{\rho}} \; = \; \frac{6}{5} \; = \; 1.2 \; .
\end{equation}
Underlying Eqs.~(2.3)--(2.6) is the CL log-divergent gap equation (LDGE)
\begin{equation}
1 \; = \; -i4N_cg^2\,\int\;\frac{d^4p}{(2\pi)^4}\,(p^2-m_q^2)^{-2} \; ,
\end{equation}
corresponding to the $V\pi\pi$ quark-loop form factors, automatically
normalized to \cite{DS95_99,PS83} $F_{\pi}\mbox{$(q^2\!=\!0)$}=1$. Further
invoking the LDGE (2.7) in turn requires \cite{DS95_99,CN77_89}
\begin{equation}
g_{\rho\pi\pi} \; = \; \sqrt{3}\,g_{\pi qq} \; = \; 2\pi \; ,
\end{equation}
close to the value $6.04$ needed to obtain the observed $\rho$ width $150.2$
MeV.

\subsection{Chiral cancellations for strong-interaction $s$-wave $\pi\pi$ and
$\pi N$ scattering lengths}
\begin{figure}[ht]
\unitlength1cm
\epsfxsize=  9.5cm
\epsfysize=  5cm
\centerline{\epsffile{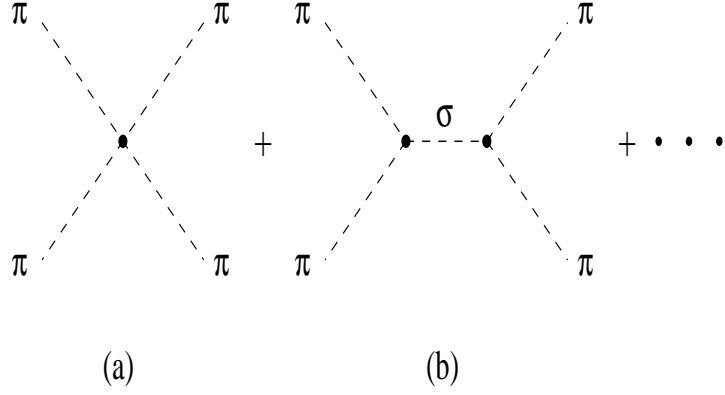}}
\caption{Low-energy $\pi\pi$ L$\sigma$M graphs.}\label{figpipi1}
\end{figure}

\begin{figure}[ht]
\unitlength1cm
\epsfxsize=  14.5cm
\epsfysize=  5cm
\vspace{1cm}
\centerline{\epsffile{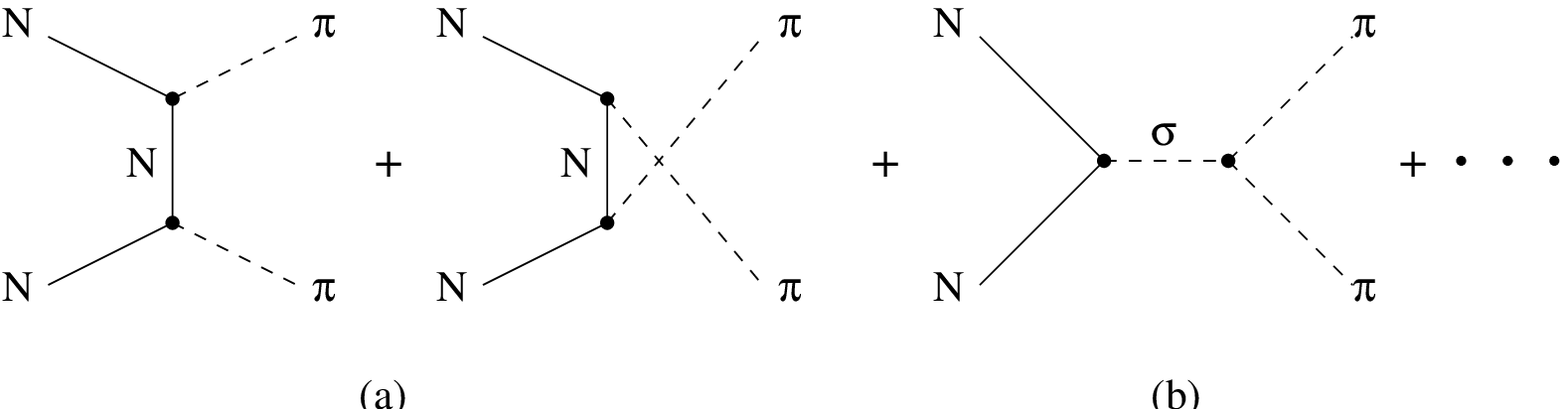}}
\caption{Low-energy $\pi N$ L$\sigma$M graphs.} \label{figpin1}
\end{figure}

Consider the low-energy $\pi\pi$ and $\pi N$ \lsm\ graphs of
Figs.~\ref{figpipi1} and \ref{figpin1}.
Away from the CL, the $\pi\pi$ contact graph with coupling $\lambda$ (Fig.~1a)
is related to the cubic meson coupling (Fig.~1b) as
\begin{equation}
g_{\sigma\pi\pi} \; ( \; = g' \; ) \; = \; \frac{m_{\sigma}^2-m_{\pi}^2}
{2f_{\pi}} \; = \;
\lambda f_{\pi} \; .
\end{equation}
Then, at the soft-pion point $s=m_{\pi}^2$, the net $\pi\pi$ amplitude
(Figs.~1a+1b) ``miraculously'' vanishes \cite{GML60}:
\begin{equation}
{\cal M}_{\pi\pi}\;=\;{\cal M}_{\pi\pi}^{\mbox{\scriptsize contact}}\;+\;
{\cal M}_{\pi\pi}^{\sigma\mbox{\scriptsize-pole}} \;\;\longrightarrow\;\;
\lambda\;+\;2g_{\sigma\pi\pi}^2(m_{\pi}^2-m_{\sigma}^2)^{-1}\;=\;0\;.
\end{equation}
In other words, the contact term $\lambda$ ``chirally eats'' the $\sigma$ pole
at $s=m^2_{\pi}$, due to Eq.~(2.9). Crossing symmetry then extends Eq.~(2.10)
to the Weinberg PCAC form \cite{W66}, but generalized to the \lsm\ \cite{S99}:
\begin{equation} \begin{array}{l} 
{\cal M}_{\pi\pi}^{abcd} \; = \; A\delta^{ab}\delta^{cd}\;+\;
B\delta^{ac}\delta^{bd}\;+\;C\delta^{ad}\delta^{bc} \; , \\[2mm] \displaystyle
A^{\mbox{\scriptsize \lsm}} \; = \; -2\lambda\left[1-\frac{2\lambda f_{\pi}^2}
{m_{\sigma}^2-s}\right] \; = \; \frac{m_{\sigma}^2-m_{\pi}^2}{m_{\sigma}^2-s}\,
\frac{s-m_{\pi}^2}{f_{\pi}^2} \; .
\end{array}\end{equation}
Thus, for $m_{\sigma}=$ 650 MeV (as dynamically generated in
Ref.~\cite{DS95_99} via the GTR), the $I=0$ $s$-channel amplitude $3A+B+C$
predicts a 23\% enhancement of the Weinberg $s$-wave scattering length at
$s=4m_{\pi}^2$, $t=u=0$, $\varepsilon=m_{\pi}^2/m_{\sigma}^2=0.045$:
\begin{equation}
\left.a_{\pi\pi}^{(0)}\right|_{\mbox{\scriptsize\lsm}}\;=\;\frac{7+\varepsilon}
{1-4\varepsilon}\,\frac{m_{\pi}}{32\pi f_{\pi}^2}\;\approx\;1.23\,
\frac{7m_{\pi}}{32\pi f_{\pi}^2}\;\approx\;0.20\,m_{\pi}^{-1} \; .
\end{equation}

If instead we use $m_{\sigma}=$ 550 MeV, a value which is closer to what is
found in unitarized meson models \cite{BR86_99,TR96}, we get $\varepsilon=$
0.063, so that Eq.~(2.12) yields an increased scattering length $a_{\pi\pi}^
{(0)}|_ {\mbox{\scriptsize\lsm}}\approx$ 0.22 $m_{\pi}^{-1}$. The latter result
is also obtained in a 2-loop chiral-perturbation-theory (ChPT) calculation
involving about 100 arbitrary LECs! So we prefer working with the simple
parameter-free \lsm\ form (2.12), since the Weinberg PCAC scattering length
\cite{W66} is based on the PCAC equation itself, first derived via the \lsm\
Lagrangian \cite{GML60}, our Eq.~(2.1).

Proceeding on to the $s$-wave $\pi N$ scattering length, the $\pi N$ background
amplitude with pseudoscalar (PS) coupling and ``Adler consistency condition''
(ACC) is \cite{A65}, for $q\rightarrow0$,
\begin{equation}
{\cal M}_{\pi N}^{ij}(\mbox{PS})\;=\;\frac{g_{\pi NN}^2}{m_N}\,\delta^{ij}\;.
\end{equation}
Then the isospin-zero scattering length corresponding to the ``large'' PS
$\pi N$ pole term, reading
\begin{equation}
a_{\pi N}^{(+)}(\mbox{PS}) \; = \; -\frac{g_{\pi NN}^2}{4\pi}\,
\frac{1}{m_N+m_\pi} \; \approx \; -1.8\,m_{\pi}^{-1} \; ,
\end{equation}
is reduced to near zero by adding to it the term of Eq.~(2.13):
\begin{equation}
a_{\pi N}^{(+)}(\mbox{Adler}) \; = \; -\frac{g_{\pi NN}^2}{4\pi}\,
\frac{m_{\pi}^2/4m_N^2}{m_N+m_\pi} \; \approx \; -0.01\,m_{\pi}^{-1} \; ,
\end{equation}
due to the ACC soft-pion theorem. Stated in \lsm\ language, when the $\sigma$
pole in Fig.~2b is added to Fig.~2a (Eq.~(2.14)), the net $\pi N$ scattering
length (due to the \lsm\ coupling (2.9)) combined with the GTR again leads to
the small scattering length (2.15) \cite{Frieder,ham67} .

These ``miraculous'' \cite{GML60} chiral cancellations, Eqs.~(2.10) and (2.15),
both due to the \lsm\ coupling (2.9), appear to follow the experimental data, 
suggesting $a_{\pi N}^{(+)}\approx-0.005\:m_{\pi}^{-1}$ back in 1979
\cite{N79},
and now finding \cite{HD01} $a_{\pi N}^{(+)}=(-0.0001\,\raisebox{-0.8mm}{$
\stackrel{+0.0009}{\scriptstyle-0.0021}$}\,)\:m_{\pi}^{-1}$
and $a_{\pi N}^{(+)}=(-0.22\pm0.43)\:m_{\pi}^{-1}$, respectively. However, ChPT
advocates prefer to work with a (seemingly non-renormalizable and obviously
non-local) pseudovector theory, derived from a non-linear $\sigma$ model, from
which the $\sigma$ meson has been eliminated as a fundamental degree of
freedom. In our opinion, this is one of the reasons why in ChPT the above
results require such a tremendous effort, while they are almost trivially
obtained in the quark-level \lsm. At this point, we can also not ignore the
mounting experimental evidence for the existence of the $\sigma$ \cite{PDG00}.

\subsection{Pion charge radius and the chiral pion}
Now we comment on the chiral structure of the pion charge radius \cite{GT79}
\begin{equation}
r^2_{\pi} \; = \; 6\,\left.\frac{dF_{\pi}(q^2)}{d(q^2)}\right|_{q^2=0} \; = \;
\frac{3}{4\pi^2f_{\pi}^2} \; = \; (0.60 \mbox{fm})^2 
\end{equation}
for the chiral-limiting value $f_{\pi}=90$ MeV, which result is close to the
measured \cite{GL84} $0.63\pm 0.01$~fm. Invoking the \lsm\ relation
\cite{DS95_99} $f_{\pi}^{\mbox{\scriptsize CL}}=\frac{\sqrt{3}}{2\pi}\,m_q$
from Eq.~(2.4) above, then Eq.~(2.16) requires, from the quark-loop pion form
factor at $q^2=0$ \cite{BRS98},
\begin{equation}
r_{\pi} \; = \; \frac{1}{m_q} \; = \; 0.61\;\mbox{fm} \; .
\end{equation}
This tightly bound (fused) pion charge radius, as observed experimentally,
certainly suggests the chiral pion's wave function is $q\bar{q}$. Note that
ChPT requires $r_{\pi}$ to be proportional to the parameter ``$L_9$''
\cite{GL84_85}. We prefer the parameter-free forms, Eqs.~(2.16) and (2.17)
above.

\subsection{Chiral couplings for $\pi^0\rightarrow2\gamma$ and
$\sigma\rightarrow2\gamma$ e.m.\ decays}
One knows that PVV \lsm\ coupling \cite{DLS99} or AVV coupling \cite{ABJ69}
gives the gauge-invariant chiral quark-loop $\pi^0\rightarrow2\gamma$ amplitude
\begin{equation}
|F_{\pi^0\rightarrow2\gamma}| \; = \; \frac{\alpha}{\pi f_{\pi}} \; \approx \;
0.025 \; \mbox{GeV}^{-1} \; ,
\end{equation}
for $N_c=3$. This is in perfect agreement with the data \cite{PDG00}
$\Gamma_{\pi^0\rightarrow\gamma\gamma}=m_{\pi}^3|F_{\pi^0\rightarrow2\gamma}|^2
/64\pi$ or $|F_{\pi^0\rightarrow2\gamma}|=0.025\pm0.001$ GeV$^{-1}$, for
$N_c=3$. Likewise, the chiral partner to the $\pi$, the $\sigma$(630), predicts
the gauge-invariant quark-loop-plus-$\pi^+$-loop amplitude \cite{KS91_93}
\begin{equation}
|F_{\sigma\rightarrow\gamma\gamma}|\;=\;\frac{5}{3}\frac{\alpha}{\pi f_{\pi}}\;
+\;0.5\frac{\alpha}{\pi f_{\pi}}\;\approx\;2.2\frac{\alpha}{\pi f_{\pi}}\;
\approx\;0.055\;\mbox{GeV}^{-1} \; ,
\end{equation}
corresponding to the decay rate (for $m_{\sigma}=$ 630 MeV)
\begin{equation}
\Gamma_{\sigma\rightarrow\gamma\gamma} \; = \; \frac{m_{\sigma}^3
|F_{\sigma\rightarrow\gamma\gamma}|^2}{64\pi} \; \approx \;3.76\;\mbox{keV} \;.
\end{equation}
This prediction is reasonably compatible with the extracted $\sigma\rightarrow
2\gamma$ rates \cite{BP99,MP90} $(3.8\pm1.5)$ keV and $(5.4\pm2.8)$ keV, 
respectively, provided that these rates indeed refer to the $\sigma$, as
advocated by the authors \cite{BP99}, and not to the \ft.

\subsection{Chiral transitions for weak $K_S\rightarrow2\pi$ decays} 
The $s$-channel $\sigma$-pole graph of Fig.~3 dominates parity-violating (PV)
$K_S\rightarrow2\pi$ decays, with PV weak amplitude magnitude
\begin{figure}[ht]
\unitlength1cm
\epsfxsize=  9.5cm
\epsfysize=  4cm
\centerline{\epsffile{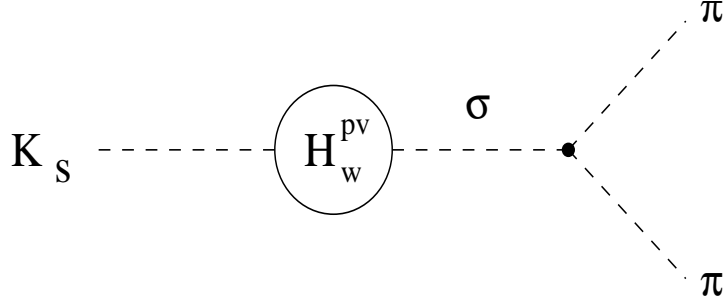}}
\caption{Parity-violating decay $K_s\rightarrow\pi\pi$ via a sigma pole.}
\label{figkspipi}
\end{figure}
\begin{equation}
\left|<2\pi|H_{w}^{\mbox{\scriptsize PV}}|K_S>\right|\;=\;|2<2\pi|\sigma>|\;
\frac{\left|<\sigma|H_{w}^{\mbox{\scriptsize PV}}|K_S>\right|}{m^2_{K_S}-
m_{\sigma}^2+im_{\sigma}\Gamma_{\sigma}}\;\approx\;\frac{1}{f_{\pi}}\:
\left|<\sigma|H_{w}^ {\mbox{\scriptsize PV}}|K_S>\right| \; ,
\end{equation}
since $<2\pi|\sigma>_{\mbox{\scriptsize \lsm}}=m_{\sigma}^2/2f_{\pi}$, 
$\:m_K\simeq m_{\sigma}$, and $\Gamma_{\sigma}\simeq m_{\sigma}$ for the broad
$\sigma$ meson \cite{PDG00}. However, pion PCAC consistency requires
\cite{KS92}
\begin{equation}
\left|<2\pi|H_{w}^{\mbox{\scriptsize PV}}|K_S>\right| \;\longrightarrow\;\frac
{1}{f_{\pi}}\:\left|<\pi|[Q_5^{\pi},H_w^{\mbox{\scriptsize PV}}]|K_S>\right|\;=
\;\frac{1}{f_{\pi}}\left|<\pi|H_{w}^{\mbox{\scriptsize PC}}|K_L>\right| \; ,
\end{equation}
for $H_w$ built up from $V\!-\!A$ chiral currents (PC = parity conserving).
Equating (2.21) to (2.22) gives a definition of chiral $\pi$ and $\sigma$
partners \cite{KS91_93}:
\begin{equation}
\left|<\sigma|H_{w}^{\mbox{\scriptsize PV}}|K_S>\right| \; = \;
\left|<\pi^0 |H_{w}^{\mbox{\scriptsize PC}}|K_L>\right| \; .
\end{equation}
The charge algebra $[Q+Q_5\,,\,H_w]\,=\,0$, PCAC, and Eq.~(2.23) clearly
suggest that the $\pi$(140) and $\sigma$(630) are chiral partners.

\section{Quark-loop \lsm\ strong and e.m.\ decays} \setcounter{equation}{0}
Rather than proceeding on with more detailed weak-interaction predictions, from
Sec.~2 we test the \lsm\ quark-loop predictions directly against the data for
strong and e.m.\ decays. First consider the $udu$ plus $dud$ quark loops for
$\rho^0\rightarrow\pi^+\pi^-$ decay with the LDGE (2.7), leading to
$g_{\rho\pi\pi}=2\pi$, Eq.~(2.8). The latter \lsm\ VMD coupling predicts the
rate
\begin{equation}
\Gamma_{\rho^0\rightarrow2\pi} \; = \; \frac{g^2_{\rho\pi\pi}}{m_{\rho}^2}\,
\frac{|{\bf p}|^3}{6\pi} \; = \; 162.6\;\mbox{MeV}\;\;\;\;\mbox{for}\;\;
{|\bf p|}=358\;\mbox{MeV} \; ,
\end{equation}
close to data at \cite{PDG00} $(150.2\pm0.8)$ MeV. For the small $\rho^0
\rightarrow e^+e^-$ and $\omega\rightarrow e^+e^-$ decays, we use single-photon
exchange to extract the $g_{\rho}$ and $g_{\omega}$ couplings from data
\cite{PDG00}:
\begin{eqnarray}
\Gamma_{\rho^0\rightarrow e^+e^-} \; = \; \frac{\alpha^2}{3}\,m_{\rho}\,
\frac{4\pi}{g_{\rho}^2} \; = \; 6.77\pm0.32\;\mbox{keV} \; , \\
\Gamma_{\omega\rightarrow e^+e^-} \; = \; \frac{\alpha^2}{3}\,m_{\omega}\,
\frac{4\pi}{g_{\omega}^2} \; = \; 0.60\pm0.02\;\mbox{keV} \; ,
\end{eqnarray}
leading to
\begin{equation}
g_{\rho} \; \approx \; 5.03 \;\;\; , \;\;\; g_{\omega} \; \approx 17.05 \; .
\end{equation}
The latter couplings are near the $U(3)$ value $g_{\omega}=3g_{\rho}$, assuming
the $\omega$ is purely non-strange. But one knows \cite{PDG00} that there is a
slight $\omega$-$\phi$ mixing angle $\phi_V\approx3.7^{^{\,\circ}}$, from the
small $\phi\rightarrow\pi\gamma$ decay. Note that the \lsm\ coupling
$g_{\rho\pi\pi}$ is relatively near $g_{\rho}\approx5.03$ found in Eq.~(3.4).
However, when one adds the $\pi$-$\sigma$-$\pi$ meson loop to the quark loop,
one knows from the \lsm\ Eq.~(2.6) that actually $g_{\rho\pi\pi}/g_{\rho}=6/5$,
whereas Eq.~(3.4) predicts the nearby ratio
\begin{equation}
\frac{g_{\rho\pi\pi}}{g_{\rho}} \; \approx \; \frac{2\pi}{5.03} \; \approx \;
1.25 \; .
\end{equation}

Next consider the e.m.\ decays $\rho\rightarrow\pi\gamma$ and
$\omega\rightarrow\pi\gamma$. Our only use of $SU(3)$ symmetry is
$\lambda_{\gamma}=\lambda_3+\lambda_8/\sqrt{3}$. This predicts the quark-loop
decays, using the $g_{\rho}$ and $g_{\omega}$ couplings from data in Eq.~(3.4),
\begin{equation}
\Gamma_{\rho\rightarrow\pi\gamma} \; = \; \frac{|{\bf p}|^{3}}{12\pi}\:
|{\cal M}_{\rho\pi\gamma}|^2 \; = \; 59 \; \mbox{keV}\;\;\;\mbox{for} \; 
|{\bf p}|=372\; \mbox{MeV} \; ,
\end{equation}
with $|{\cal M}_{\rho\pi\gamma}|=eg_{\rho}/8\pi^2f_{\pi}=0.207$ GeV$^{-1}$,
which comes out close to the data \cite{PDG00} $\Gamma_{\rho^{\pm}\rightarrow
\pi^{\pm}\gamma}=(68\pm7)$ keV. Likewise, the \lsm\ predicts
\begin{equation}
\Gamma_{\omega\rightarrow\pi\gamma} \; = \; \frac{|{\bf p}|^{3}}{12\pi}\:
|{\cal M}_{\omega\pi\gamma}|^2 \; = \; 711 \; \mbox{keV}\;\;\;\mbox{for} \; 
|{\bf p}|=379\; \mbox{MeV} \; ,
\end{equation}
with $|{\cal M}_{\omega\pi\gamma}|=eg_{\omega}\cos{\phi_V}/8\pi^2f_{\pi}=
0.7017$ GeV$^{-1}$, very close to the data \cite{PDG00} $\Gamma_{\omega
\rightarrow\pi\gamma}=(717\pm43)$ keV. 

Finally, the e.m.\ decay $\pi^0\rightarrow2\gamma$ is predicted via $u$ and $d$
quark loops, together with the gauge-invariant amplitude (2.18) and $N_c=3$, to
be
\begin{equation}
\Gamma_{\pi^0\rightarrow2\gamma} \; = \; \frac{|{\bf p}|^{3}}{8\pi}\:
|{\cal M}_{\pi^02\gamma}|^2 \; = \; 7.64 \; \mbox{eV}\;\;\;\mbox{for} \; 
|{\bf p}|=67.49\;\mbox{MeV} \; ,
\end{equation}
again close to the data \cite{PDG00} $\Gamma_{\pi^0\rightarrow2\gamma} =
(7.74\pm0.55)$ eV.

When considering $\eta$ and $\eta^{\prime}$ initial states, we circumvent
explicit $\eta$-$\eta^{\prime}$ mixing by only computing the sum of their
squared matrix elements, thereby using $\cos^2{\!\phi_{PS}}+\sin^2{\!\phi_{PS}}
=1$. Then, Table~I of Ref.~\cite{DLS99} shows the PVV quark-loop matrix
elements for $\pi^0\rightarrow2\gamma$, $\eta\rightarrow2\gamma$,
$\eta^{\prime}\rightarrow2\gamma$ are, respectively, $A$, $A(5\cos{\phi_{PS}}-
\sqrt{2}r_s\sin{\phi_{PS}})/3$, $A(5\sin{\phi_{PS}}+\sqrt{2}r_s\cos{\phi_{PS}})
/3$, for $A=\alpha/\pi f_{\pi}\approx 0.025$ GeV$^{-1}$, and where
$r_s=\hat{m}/m_s\approx1/1.44$ is the constituent-quark-mass ratio,
with $m_s/\hat{m}=2f_K/f_{\pi}-1$ and $f_K/f_{\pi}=1.22$. Therefore, the
matrix-element squares satisfy $|{\cal M}_{\eta2\gamma}|^2+
|{\cal M}_{\eta^{\prime}2\gamma}|^2=A^2(25+2/1.44^2)/9$, corresponding to the
\lsm\ decay-rate $SU(3)$ \/sum rule, implied from Ref.~\cite{DLS99},
\begin{equation}
\frac{\Gamma_{\eta2\gamma}}{m_{\eta}^3} \; + \;
\frac{\Gamma_{\eta^{\prime}2\gamma}}{m_{\eta^{\prime}}^3} \; = \; 2.885 \:
\frac{\Gamma_{\pi^02\gamma}}{m^3_{\pi^0}} \; .
\end{equation}
Given the measured central-value rates and masses \cite{PDG00}
$\Gamma_{\eta2\gamma}=464$ eV, $m_{\eta}=0.5473$ GeV,
$\Gamma_{\eta^{\prime}2\gamma}=4282$ eV, $m_{\eta^{\prime}}=0.9578$ GeV,
$\Gamma_{\pi^02\gamma}=7.74$ eV, $m_{\pi^0}=0.1349766$ GeV, the l.h.s.\ of
Eq.~(3.9) sums to $7704\times10^{-9}$ GeV$^{-2}$, while the r.h.s.\ is
$9081\times
10^{-9}$ GeV$^{-2}$.  A one-standard-deviation reduction of the r.h.s.\ gives
$8435\times10^{-9}$ GeV$^{-2}$, only 9\% greater than the l.h.s.\ of Eq.~(3.9).

Likewise, we can construct an $SU(3)$ sum rule again by invoking
$\sin^2{\!\phi}+\cos^2{\!\phi}=1$ for any angle, and referring to
Ref.~\cite{DLS99} for $\eta^{\prime}\rightarrow\rho\gamma$, $\rho\rightarrow
\eta\gamma$, $\rho\rightarrow\pi\gamma$ decays. The squares of the \lsm\
quark-loop matrix elements are 
$|{\cal M}_{\eta^{\prime}\rho\gamma}|^2+|{\cal M}_{\rho\eta\gamma}|^2=9B^2$ and
$|{\cal M}_{\rho\pi\gamma}|^2=B^2$, corresponding to the \lsm\ decay-rate
$SU(3)$ sum rule \cite{DLS99}
\begin{equation}
\frac{\Gamma_{\eta^{\prime}\rho\gamma}}{|{\bf p}_1|^3} 
\;+\; 3\:\frac{\Gamma_{\rho\eta\gamma}}{|{\bf p}_2|^3} 
\;=\;27\:\frac{\Gamma_{\rho\pi\gamma}}{|{\bf p}_3|^3} \; .
\end{equation} 
For the measured central-value rates and CM momenta \cite{PDG00}
$\Gamma_{\eta^{\prime}\rho\gamma}=59.6$ keV, $|{\bf p}_1|=169$ MeV;
$\Gamma_{\rho\eta\gamma}=36.05$ keV, $|{\bf p}_2|=189$ MeV;
$\Gamma_{\rho\pi\gamma}=67.6$ keV, $|{\bf p}_3|=372$ MeV, the l.h.s.\ of the
sum rule Eq.~(3.10) sums up to $1.2348\times10^{-2}$ GeV$^{-2}\:+\:1.6019
\times10^{-2}$ GeV$^{-2} = 2.8367\times10^{-2}$ GeV$^{-2}$, while the r.h.s.\
is $3.5455\times10^{-2}$ GeV$^{-2}$. Considering we have combined the PS
$\eta^{\prime}$ decay rate and two vector $\rho$ decay rates, we suggest that
the \lsm\ $SU(3)$ sum rule (3.10) is reasonably well satisfied.

By analogy with Eqs.~(3.9) and (3.10), another $SU(3)$ sum rule implied in
Ref.~\cite{DLS99} reads
\begin{equation}
\frac{\Gamma_{\eta^{\prime}\omega\gamma}}{|{\bf p}_a|^3} 
\;+\; 3\:\frac{\Gamma_{\omega\eta\gamma}}{|{\bf p}_b|^3} \;=\;0.336\:
          \frac{\Gamma_{\omega\pi\gamma}}{|{\bf p}_c|^3} \; .
\end{equation}
From Ref.~\cite{PDG00}, the measured central-value rates and CM momenta are
$\Gamma_{\eta^{\prime}\omega\gamma}=6.12$ keV, $|{\bf p}_a|=160$ MeV;
$\Gamma_{\omega\eta\gamma}=5.486$ keV, $|{\bf p}_b|=199$ MeV;
$\Gamma_{\omega\pi\gamma}=717$ keV, $|{\bf p}_c|=379$ MeV. Then the l.h.s.\ of
Eq.~(3.11) sums to $1.4941\times10^{-3}$ GeV$^{-2}\:+\:2.0884\times10^{-3}$
GeV$^{-2} = 3.5825\times10^{-3}$ GeV$^{-2}$, while the r.h.s.\ is nearby at
$4.4253\times10^{-3}$ GeV$^{-2}$. If one increases the
$\omega\rightarrow\eta\gamma$ rate by one standard deviation, the l.h.s.\ of
Eq.~(3.11) becomes $3.940\times10^{-3}$ GeV$^{-2}$ --- again only 9\% below the
r.h.s.~.

\section{Conclusions}
In the preceding we have shown, by straightforward computation, that the
quark-level \lsm\ of Refs.~\cite{S82_92,DS95_99} easily reproduces the small
$\pi\pi$ and $\pi N$ $s$-wave scattering lengths, the pion charge radius, and
a variety of e.m., weak, and strong decays of pseudoscalar and vector mesons.
Crucial for most of these processes is the inclusion of the $f_0$(400--1200),
alias $\sigma$ meson, as a fundamental $q\bar{q}$ degree of freedom. This
occurs very naturally in the \lsm, where the $\sigma$ can then also be
dynamically and self-consistently generated \cite{DS95_99}, as well as in the
unitarized quark/meson model of Ref.~\cite{BR86_99}. Moreover, a
finite-temperature (recall Ref.~\cite{Frieder}) chiral-phase-transition
approach \cite{BBSZ85_01}, which
independently ``melts'' the quark mass, the $\sigma$ mass, and the quark
condensate in QCD, suggests that the above \lsm\ can be identified as the
infrared limit of QCD \cite{BES97}.

In contrast, non-linear
approaches where the $\sigma$ does not show up as a $q\bar{q}$ state or has
even been designedly eliminated as a fundamental degree of freedom, like the
confining NJL-type model of SW \cite{SW_prd64,SW_prd63} and ChPT, appear to
have difficulties in reproducing several low-energy data, besides having
strained relations with the now firmly established $\sigma$ itself
\cite{GL84_85,KS91_93}.
We therefore argue that the conclusion of SW \cite{SW_prd64} according to which
the $\sigma$ is \em not \em \/the chiral partner of the pion is not based on
\em ``major chiral-symmetry violations'', \em but rather on the complications
and possible approximations in their non-linear NJL scheme. In this respect, we
should point out the following apparent contradiction in SW's line of
reasoning. In Ref.~\cite{SW_prd64} they conclude that confinement is quite a
small effect for the $\pi$(138) and $K$(495) mesons, which may even be best to
neglect altogether. However, in Sec.~2.3 we showed that the observed pion
charge radius suggests in fact a $q\bar{q}$ (fused) $\pi$ meson composed of
tightly bound quarks, corresponding to an almost massless Nambu-Goldstone pion.
Moreover, the well-understood NJL model \em without \em
\/confinement \em does \em \/predict a bound-state $\sigma$ meson as the chiral
$q\bar{q}$ partner of the pion. We believe to have demonstrated, in the
framework of the quark-level \lsm, that this is indeed the scenario favored by
experiment.  \\[1cm]
{\it Acknowledgements.} \\[1mm]
The authors wish to thank D.~V.~Bugg for very useful discussions on processes
involving scalar mesons. 
This work was partly supported by the
{\em Funda\c{c}\~{a}o para a Ci\^{e}ncia e a Tecnologia} (FCT) 
of the {\em Minist\'{e}rio da Ci\^{e}ncia e da Tecnologia} of 
Portugal, under Grant no.\ PRAXIS XXI/\-BPD/\-20186/\-99 and under contract
numbers POCTI/\-35304/\-FIS/\-2000 and CERN/\-P/\-FIS/\-40119/\-2000.

\end{document}